\documentstyle[prb,aps,multicol,epsf]{revtex}
\catcode`\"=\active \let"=\" \let\3=\ss 
\newcommand{\Kr}[1]{\left( #1\right)}
\renewcommand{\d}{\displaystyle}
\newcommand{\ve}{\varepsilon}
\newcommand{\bm}[1]{\mbox{\boldmath$#1$}}
\newcommand{\vf}{v_{\rm F}}
\newcommand{\kf}{k_{\rm F}}
\newcommand{\ef}{\epsilon_{\rm F}}
\begin{document}
\draft
\title{Dephasing in Rashba spin precession along
mutlichannel quantum wires and nanotubes}
\author{Wolfgang H\"ausler$^*$}
\address{Physikalisches Institut,
Albert-Ludwigs-Universit\"at,
Hermann-Herder-Stra\3e 3,\\
79104 Freiburg, Germany}
\maketitle
\begin{abstract}
Coherent Rashba spin precession along interacting multi-mode
quantum channels is investigated, revisiting the theory of
coupled Tomonaga-Luttinger liquids. We identify
susceptibilities as the key-parameters to govern exponents
and Rashba precession lengths. In semiconducting quantum
wires spins of different transport channels are found to {\em
dephase\/} in their respective precession angles with respect to
one another, as a result of the interaction. This could explain
the experimental difficulty to realize the Datta Das transistor.
In single walled carbon nanotubes, on the other hand,
interactions are predicted to suppress dephasing between the two
flavor modes at small doping.
\end{abstract}

\pacs{PACS numbers: 71.10.Pm; 71.70.Ej; 72.25.Dc; 73.21.Hb; 73.22.-f}
\begin{multicols}{2}%
\narrowtext%
\section{Introduction}
Rashba precession \cite{rashba} of spins and its manipulation
has been studied intensively \cite{rashbamod} in recent years,
aiming to control the coherent propagation of electron spins.
One goal is to realize a spin transistor \cite{dattadas};
another interesting option could be to switch between singlet
and triplet entangled states for directing, for example, noise
statistics \cite{egues} or for quantum computing. Despite of
major efforts, manipulation of coherent Rashba precession could
not be demonstrated experimentally yet in the
polarizer--analyzer type transport arrangement. Rashba spin
splitting occurs near surfaces in the presence of internal or
externally applied symmetry breaking electric fields
perpendicular to the transport direction \cite{rashba}. Spins
precess when injected out of the spin-orbit eigendirections.
Any dephasing along the structure limits successful transistor
operation. One important dephasing mechanism in spin-orbit
active structures of more than one dimension arises due to
momentum randomizing scattering events by impurities
(elastically), by phonons (inelastically) \cite{DP}, or by
electron-electron scattering events \cite{wu}. One-dimensional
structures confine the direction of propagation, thus reducing
this source of dephasing mechanism \cite{pramanik}. Already in
their original proposal \cite{dattadas} Datta and Das therefore
suggested to use clean one-dimensional structures for the
spin-orbit active medium. The current work focuses on quantum
wires and assumes absence of momentum randomizing scattering
events. No spin {\em relaxing\/} mechanisms (in the sense that
off-diagonal entries of the spin density matrix decay) will be
considered here.

Most quantum wires of current experiments accommodate more than
one transport channel at the Fermi energy. This holds true also
for single walled carbon nanotubes (NT) where at least two
flavor channels carry (spin) current. In NT spin transport has
been established experimentally \cite{alphenaar} and studied
theoretically \cite{balentsegger}. Other experimental
multichannel systems consist of arrays of quantum wires
fabricated artificially in parallel at close proximity
\cite{yacoby02}. In any multichannel wire the interesting
question arises whether Rashba spin phases increase by equal
amounts along different channels or not. Provided the kinetic
energy dispersion is strictly parabolic, described by a common
effective carrier mass in each channel as in most semiconducting
quantum wires, one would expect equal spin phases in all
channels as a result of the linear splitting by the Rashba
energy \cite{dattadas}, so that their probabilities to enter
the spin selective drain contact all add up. We shall
demonstrate that Coulomb interactions between charged electrons
affect their spin propagation properties and ultimately destroy
the phase relationship of Rashba precessing spins between
channels in quantum wires (QW) fabricated on the basis of
semiconducting material. This limits operation of the Datta Das
spin transistor. In metallic single walled nanotubes, on the
other hand, with their linear kinetic energy dispersion, spins
dephase between modes already without accounting for
interactions; remarkably, in this case we find that the Coulomb
interaction {\em suppresses\/} this single particle dephasing
between the two flavor modes, particularly at small doping, thus
facilitating coherent spin precession along NT.

\section{Interacting multimode quantum wires}
Before addressing the effect of spin-orbit coupling let us first
discuss the (possibly screened) Coulomb interactions within and
between transport channels. One-dimensional systems are
particularly susceptible to electron-electron interactions, even
when weak \cite{larkin}. Neither in semiconducting QW
\cite{yacoby} nor in NT \cite{bockrath,eg} interactions can be
disregarded. Contrary to higher dimensions they show up, for
example, as nonuniversal power laws at low energies. The most
convenient theoretical framework is the Bosonic
Tomonaga--Luttinger liquid (TL) \cite{voit} developed since the
seminal work by Haldane \cite{haldane}. However, {\em a
priori}, it is not clear in how far the TL-model can be applied
to multichannel situations. Strictly equivalent modes of equal
particle densities and interaction matrix elements, for example,
tend to stabilize a gapped charge or spin density, non-TL low
energy phase \cite{finkelstein} due to the appearance of
relevant (in the sense of a perturbative renormalization group
treatment) momentum conserving intermode backscattering
processes. On the other hand, most real systems lack such a
strict equivalence of modes. In multichannel quantum wires
particle densities differ at given Fermi energy, and even
systems of Ref.~\onlinecite{yacoby02} without fine tuning of
densities, or the two flavor modes of NT at not exact zero
doping cease to be strictly equivalent. Here, we therefore focus
on the generic case of unequal modes which are expected to stay
in the gap-less low energy TL phase.

In this phase Matveev and Glazman (MG) have computed a power law
for the density of states of spin-less electrons at the end of
the $n$-th mode \cite{mg}:
\begin{equation}\label{exponent}
\nu_n(\omega)\sim\omega^{\beta_n}\;,
\end{equation}
generalizing the single channel case
\cite{larkin,haldane,kanefisher}. MG obtained the exponents
$\beta_n$ within a pure plasmon model by modeling the charge
density fluctuations as coupled harmonic strings.
Plasmon velocities $s_{\ell}$ and the normalized eigenmodes
$\gamma_{n\ell}$, as obtained from the dynamical matrix,
determine \cite{mg}
\begin{equation}\label{betamg}
\beta_n^{\mbox{\rm\tiny
MG}}=-1+\sum_{\ell}|\gamma_{n\ell}|^2s_{\ell}/v_n\;,
\end{equation}
$v_n$ is the Fermi velocity in the $n$-th channel. This
approach merely accounts for long wave length charge properties
and can shown to be equivalent to the random phase approximation
(RPA) to plasmon velocities \cite{joynt} generalized to coupled
modes \cite{unpublished}. It has been pointed out
\cite{whlkahm} that the RPA result deserves improvement at small
particle densities. Moreover, it disregards exchange processes
and spin, and tacitly presumes `super' Galilei invariance of
every channel individually, as discussed below. Spin properties
and exchange are known to depend on short wave length properties
of the interaction \cite{calmels}.

For given microscopic interaction the parameters of the single
channel TL model, and therewith sound velocities and exponents,
have been obtained from homogeneous and static susceptibilities
\cite{whlkahm}, exploiting {\em exact\/} thermodynamic relations
\cite{haldane,voit,schulz90}. These susceptibilities, in turn,
can be computed to high accuracy by standard many-body
techniques from the underlying Fermion model, beyond the RPA or
perturbative accuracies. This way the asymptotic behavior of
correlation functions has been determined in the 1D Hubbard
model from the Bethe Ansatz ground state energy $E_0$
\cite{schulz90}. In the Galilei invariant charge sector of QW
only the compressibility
$\kappa=\Kr{L\frac{\partial^2E_0}{\partial N^2}}^{-1}$ is
required ($N$ is the particle number and $L$ the system length)
to fix the exponent parameter $K_{\rho}=\sqrt{\pi\kappa\vf/2}$.
In spin sector SU(2) spin rotation invariance enforces $K_{\sigma}=1$
(cf., e.g., Refs.~\onlinecite{voit} and \onlinecite{giamarchibook}).
As a further important property of one dimension charge and spin
density wave excitations are expected to separate and move at
different velocities \cite{haldane}. Evidence for this has
been found in recent experiment \cite{yacoby02}. Spin
velocities of QW have been deduced from quantum Monte Carlo
magnetic susceptibilities \cite{creffield}. Here we generalize
this {\em a priori\/} exact thermodynamic approach to coupled
channels. Spins can be incorporated as separate $s=\pm$ modes
which allows to account even for non SU(2) invariant situations,
as it arises for example in the presence of a Zeeman-field
\cite{aoki}.

From the microscopic point of view we consider the 1D system
\begin{eqnarray}\label{model}
&&H=\sum_{n,k,s}\epsilon_n(k)c_{n,k,s}^\dagger c_{n,k,s}^{}+
\frac{1}{2L}\times\\
&&\d\sum_{k,s,k',s',q}c_{n_1,k-q,s}^\dagger c_{n_2,k'+q,s'}^\dagger
V_{n_1n_2n_3n_4}(q)c_{n_3,k',s'}^{}c_{n_4,k,s}^{}\;.\nonumber
\end{eqnarray}
Fermi annihilation operators $c_{n,k,s}$ refer to wave vector
$k$ and spin $s$ of mode $n$. Assuming electron wave functions
of the product form $\:\sim\varphi_n(x_{\perp}){\rm e}^{{\rm
i}kx}/\sqrt{L}\:$ (for a discussion of this product assumption
cf.\ Ref.~\onlinecite{whlkahm}) we obtain
\begin{eqnarray}\label{interaction}
&&\d V_{n_1n_2n_3n_4}(q)=\frac{2e^2}{\ve}\int{\rm d}x_{\perp}
\int{\rm d}x_{\perp}'\times\\
&&\d\varphi_{n_1}^*(x_{\perp})\varphi_{n_2}^*(x_{\perp}')
\varphi_{n_3}^{}(x_{\perp}')\varphi_{n_4}^{}(x_{\perp})
{\rm K}_0(|q||x_{\perp}-x_{\perp}'|)
\nonumber
\end{eqnarray}
from the 3D-Coulomb interaction between electrons, $\ve$ is the
dielectric constant of the material surrounding the wire. The
$V_{n_1n_2n_3n_4}$ can be expressed analytically for many cross
sections of physical relevance \cite{oconnel,glazman}. Some of
the $V_{n_1n_2n_3n_4}$ vanish by symmetry: for example, angular
momentum conservation on a cylinder surface of NT requires
$n_1+n_2=n_3+n_4$, or $\sum_in_i$ must be even in QW of mirror
symmetric cross section. `Direct' terms may be approximated as
$V_{nn'n'n}^{\mbox{\tiny QW}}(q)\approx\frac{e^2}{\ve} {\rm
e}^{\tilde q}{\rm K}_0(\tilde q)$ or $V_{nn'n'n}^{\mbox{\tiny
NT}}(q)\approx\frac{2e^2}{\ve}{\rm I}_0(r|q|){\rm K}_0(r|q|)$ at
$q^{-1}$ larger than the diameter (QW) $d$ or the radius $r$
(NT). At small $q$ both reveal the same logarithmic increase
which eventually will be screened by remote metallic gates. In
the above formulae I$_\nu$ and K$_\nu$ are Bessel functions and
$\tilde q=d^2q^2/8$. `Exchange' terms between the lowest two
modes of parabolically confined QW are $V_{1212}^{\mbox{\tiny
QW}}(q)=\frac{e^2}{\ve}\tilde q{\rm e}^{\tilde q}({\rm
K}_1(\tilde q)-{\rm K}_0(\tilde q))$ \cite{oconnel} while in NT
the two lowest degenerate flavor modes have angular momentum
zero and interact $\sim V_{1111}^{\mbox{\tiny NT}}(q)$.

The Boson model \cite{haldane}, describing gapless excitations
of Fermions in a 1D wire of length $L$ (periodic boundary
conditions) in the vicinity of the Fermi energy, takes the form
\begin{equation}\label{mbose}
H=\sum_{q\ne 0}\Kr{H_q^{(1)}+H_q^{(2)}}+\frac{\pi}{4L}
\Kr{\vec{N}\bm{v}_{\mbox{\tiny N}}\vec{N}+
\vec{J}\bm{v}_{\mbox{\tiny J}}\vec{J}}
\end{equation}
when generalized to $M$ modes. Bold letters indicate $M\times
M$ matrices and the components of vectors refer to modes $1\le
n\le M$. In Eq.~(\ref{mbose})
\begin{eqnarray}
H_q^{(1)}&=&\d\frac{\pi}{L}\sum_{r=\pm}\vec\varrho_r(q)
(\bm{v}+\frac{\bm{V}^{(1)}}{\pi})\vec\varrho_r(-q)\nonumber \\
H_q^{(2)}&=&\d\frac{\pi}{L}\sum_{r=\pm}\vec\varrho_r(q)
(\frac{\bm{V}^{(2)}}{\pi})\vec\varrho_{-r}(-q)
\label{hq}
\end{eqnarray}
describe excitations of right/left ($r=+/-$) going Bosonic
density fluctuations $\varrho_{rn}(q)$ at wave numbers $q\ne 0$.
Topological excitations [last term in Eq.~(\ref{mbose})] describe
changes in the ground state energy when particles $N_n$ or
currents $J_n$ are added to mode $n$ at $q=0$. This second part
of Eq.~(\ref{mbose}) is important in the present context and is
governed by generalized homogeneous and static compressibilities
and (Drude) conductivities:
\begin{equation}\label{compresss}
(\bm{v}_{\mbox{\tiny N}})_{nn'}=\frac{2L}{\pi}\frac{\partial^2E_0}
{\partial N_n\partial N_{n'}}\quad\mbox{and}\quad
(\bm{v}_{\mbox{\tiny J}})_{nn'}=\frac{2L}{\pi}\frac{\partial^2E_0}
{\partial J_n\partial J_{n'}}\;,
\end{equation}
respectively. They generalize corresponding TL parameters for
single channels \cite{voit} ($E_0$ is the ground state energy of
the interacting electron system) and govern the complete low
energy physics. As in the single channel case \cite{haldane}
$\bm{v}_{\mbox{\tiny N}}$ and $\bm{v}_{\mbox{\tiny J}}$ are
related to the interactions between density fluctuations in
Eq.~(\ref{hq}), moving in the same
\begin{equation}\label{v1}
\bm{V}^{(1)}/\pi=(\bm{v}_{\mbox{\tiny N}}+
\bm{v}_{\mbox{\tiny J}})/2-\bm{v}
\end{equation}
or opposite
\begin{equation}\label{v2}
\bm{V}^{(2)}/\pi=(\bm{v}_{\mbox{\tiny N}}-\bm{v}_{\mbox{\tiny J}})/2
\end{equation}
directions, assuming $\epsilon_n(k)=\epsilon_n(-k)$ and thus
Fermi velocities $(\bm{v})_{nn'}=v_n\delta_{nn'}$ of equal
magnitudes at either Fermi point.

Importantly, all entries of $\bm{v}_{\mbox{\tiny N}}$ and
$\bm{v}_{\mbox{\tiny J}}$ are observable, at least {\em in
principle\/}, and must therefore agree in Bosonic (\ref{mbose})
or Fermionic [Eq.~(\ref{model}) together with (\ref{compresss})]
representation. This allows deduction $\bm{V}^{(1)}$ and
$\bm{V}^{(2)}$ by standard many-body techniques from the given
microscopic interaction (\ref{interaction}) through
Eqs.~(\ref{compresss}---\ref{v2}). Therefore $\bm{v}_{\mbox{\tiny
N}}$ and $\bm{v}_{\mbox{\tiny J}}$ (and not exponents as often
assumed for single channel quantum wires) are the principle
parameters governing the low energy physics in the multichannel
case. They establish the quantitative link to the microscopic
Fermion model (\ref{model}). For real quantum wires
self-consistent Hartree-Fock \cite{whlkahm,whahm}, diagrammatic
\cite{calmels}, or Monte Carlo techniques \cite{creffield} were
used to deduce TL parameters from a microscopic model for
electron-electron interaction potential. As a first approach to
multichannels we rely on the perturbative approximation below,
which, when including the Fock term (as crucial in spin sector),
proves already as superior to the often used random phase
approximation \cite{joynt} to TL parameters \cite{whlkahm}.

The matrices $\bm{v}_{\mbox{\tiny N}}$ and $\bm{v}_{\mbox{\tiny
J}}$ reflect symmetries of the system under permutations of
modes. One important case are equivalent modes when
$(\bm{v})_{nn'}=v\delta_{nn'}$ in Eq.~(\ref{hq}), and
$\bm{v}_{\mbox{\tiny N}}$ and $\bm{v}_{\mbox{\tiny J}}$ both are
cyclic matrices. Then Eq.~(\ref{mbose}) can be diagonalized and
the Bogoliubov transformation solving for plasmon velocities and
exponents can be carried out separately in each of the resulting
independent `normal mode' TL (two equivalent channels, for
example, can be separated trivially into independent
$\varrho_1+\varrho_2$ and $\varrho_1-\varrho_2$ modes). Another
symmetry is Galilei invariance (observed in QW and NT) for which
${\rm Tr}\,\bm{v}_{\mbox{\tiny J}}=\sum_nv_n$ \cite{conductqu}
stays independent of interactions. The higher symmetry, where
the ground state energy changes only by the trivial kinetic part,
independently of the interaction strength, under boosting any
individual channel, $J_n\to J_n+\delta_n$, we call `super' Galilei
invariance. It implies $\bm{v}_{\mbox{\tiny J}}=\bm{v}$, i.e.,
$\bm{V}^{(1)}=\bm{V}^{(2)}$ which in general is not observed by
Eq.~(\ref{model}) though tacitly assumed often in theoretical work.

Of central importance is the single particle density matrix
$\langle\psi_n^{}(x)\psi_{n'}^{\dagger}(0)\rangle$ which can be
evaluated from Eq.~(\ref{mbose}) using Boson operators
$a_{nq}=\sqrt{2\pi/L|q|}\sum_r\Theta(rq)\varrho_{rn}(q)$ with
$[a_{nq}\,,\,a_{n'q'}^{\dagger}]=\delta_{nn'}\delta_{qq'}$,
generalizing the single channel case \cite{haldane,voit}.
New Boson operators $b_{nq}$ eventually diagonalize
$H_q^{(1)}+H_q^{(2)}$. They are obtained via a Bogoliubov
transformation reading in the multicomponent case $\Kr{\vec
a^{\dagger}\atop\vec a}=\Kr{{{\bf u}\;{\bf v}\atop{\bf v}\;{\bf
u}}}\Kr{\vec b^{\dagger}\atop\vec b}$. Here, the $M\times M$
matrices {\bf u} and {\bf v} must satisfy $\:({\bf w}[{\bf
vuv}^{-1}({\bf wvuv}^{-1})^{\sf t}-{\bf v}({\bf wv})^{\sf
t}])_{ll'}=\delta_{ll'}\:$ where ${\bf w}=({\bf vuv}^{-1}{\bf
u}-{\bf v}^2)^{-1}$ and {\sf t} indicates the transpose. This
condition is fulfilled when
\begin{equation}\label{bogo1}
{\bf u}={\bf R}\;{\bf c}\;{\bf R}\qquad\mbox{and}\qquad
{\bf v}={\bf R}\;{\bf s}\;{\bf R}\;,
\end{equation}
provided {\bf R} is orthogonal, and {\bf
c}$_{ll'}=\delta_{ll'}\cosh\vartheta_{l}$ and {\bf
s}$_{ll'}=\delta_{ll'}\sinh\vartheta_{l}$ are diagonal matrices,
$1\le l,l'\le M$. The $\frac{1}{2}M(M-1)$ real parameters of
{\bf R} and the $M$ angles $\vartheta_{l}$ are found from the
$\frac{1}{2}M(M+1)$ equations
\begin{equation}\label{bogo2}
\raisebox{2ex}{$\d(\matrix{{\bf c}&{\bf s}})$}
\Kr{\matrix{\bm{\tilde V}^{(2)}/\pi&\bm{\tilde v}+
\bm{\tilde V}^{(1)}/\pi\cr
\bm{\tilde v}+\bm{\tilde V}^{(1)}/\pi&\bm{\tilde V}^{(2)}/\pi}}
\Kr{\matrix{{\bf c}\cr{\bf s}}}=0\;.
\end{equation}
Here, the tilde connotes rotated symmetric matrices, $\bm{\tilde
A}={\bf R}^{-1}\bm{A}{\bf R}=\bm{\tilde A^{\sf t}}$.

Finally, the $M$ eigenvalues of
\begin{equation}\label{eigenv}
\raisebox{2ex}{$\d(\matrix{{\bf c}&{\bf s}})$}
\Kr{\matrix{\bm{\tilde v}+\bm{\tilde V}^{(1)}/\pi&\bm{\tilde V}^{(2)}/\pi\cr
\bm{\tilde V}^{(2)}/\pi&\bm{\tilde v}+\bm{\tilde V}^{(1)}/\pi}}
\Kr{\matrix{{\bf c}\cr{\bf s}}}\;,
\end{equation}
with the outcome of Eq.~(\ref{bogo2}) inserted for {\bf c}, {\bf s},
and {\bf R}, are the (generalized) plasmon velocities of
$\sum_{q\ne 0}\Kr{H_q^{(1)}+H_q^{(2)}}$. The same sequence of
transformations does not, in general, diagonalize the
topological excitations which explains why coupled TLs cannot
necessarily be decomposed into independent `normal mode TLs'.

Equations~(7---12) allow to calculate the (asymptotic) power law
behavior of any Fermion function, cf.\ Eq.~(\ref{betann}) below.
This completes the solution for low energy properties in
multi-mode electron liquids in 1D. In the (super--Galilei
invariant) special case,
$(\bm{V}^{(1)})_{nn'}=(\bm{V}^{(2)})_{nn'}=V_0$ for all $n$ and
$n'$, the ensuing open boundary exponent as well as the plasmon
velocities Eq.~(\ref{eigenv}) agree with the results of
Ref.~\onlinecite{mg}. As stated above, the present approach
includes spins through separate modes $s=\pm 1$.

With SU(2) spin rotation invariance charge-spin separation
continues to occur in multichannel systems, so that introducing
the additional index $\nu=\rho,\sigma$ for the charge and spin
sector, respectively, renders $\bm{v}_{\mbox{\tiny N}}$ and
$\bm{v}_{\mbox{\tiny J}}$ block diagonal. Furthermore, the
chiral SU(2)$\times$SU(2) symmetry of the fermionic model at low
energies provides separate spin rotation invariance of right and
left movers, giving rise to $\bm{v}_{\mbox{\tiny
N}\sigma}=\bm{v}_{\mbox{\tiny J}\sigma}$ in spin sector. Very
strong spin-orbit coupling will spoil charge-spin separation in
general. In QW, however, charge and spin are mixed only to the
order ${\cal O}(\alpha^5)$ \cite{rashbawh} which therefore is
rarely expected to be of importance [cf.\ Eq.~(\ref{hso}) below,
$\alpha$ is the spin-orbit coupling parameter]. In NT, on the
other hand, charge-spin separation is already broken to the
order ${\cal O}(\alpha)$. Here, the very small value of
$\alpha$, expected from the small carbon mass, leaves this
breaking less important and justifies calculating Rashba
precession lengths to the leading order in $\alpha$. With
charge-spin separation Rashba precession is controlled solely by
the spin sector $\nu=\sigma$ where, from now on, the index $n$
runs only over $M$ spatial modes.

\section{Rashba spin precession length}
We address the physical situations referred to in the Introduction
by studying two model systems: QW for quantum wires and NT for
nanotubes. In QW the single particle kinetic energy dispersion
may be regarded as $\epsilon_n^{\mbox{\tiny
QW}}(k)=\epsilon_n+k^2/2m_{\rm e}$ where $\epsilon_n$ are
subband energies and $m_{\rm e}$ is the effective mass. The
Rashba spin-orbit energy
\begin{equation}\label{hso}
H^{\rm so}=\alpha(\sigma_xk_z-\sigma_zk)\;,
\end{equation}
deemed as independent of the mode index, adds a linear contribution
$\pm\alpha k$ and splits $\epsilon_n^{\mbox{\tiny QW}}(k)$ into
two equal parabolas intersecting at the origin (this statement
holds strictly true only for not unrealistically strong spin
orbit coupling strengths, $\alpha\ll v_n$, as discussed above
\cite{rashbawh,lt23wh}). Both branches are indexed by the electron
spin projection perpendicular to the wire axes, taken as the
$z$-direction for convenience (the $x$-axes points along the
wire). Within each mode, at the Fermi energy, the spin
splitting yields a finite difference of Fermi momenta
$k_{n{\tiny\uparrow}}-k_{n{\tiny\downarrow}}=2\alpha m_{\rm e}$
which, as important property of the QW case, does not depend on
$n$. Spins initially polarized along the $x$-axes therefore
precess once over a length
$4\pi/|k_{n{\tiny\uparrow}}-k_{n{\tiny\downarrow}}|=2\pi/\alpha
m_{\rm e}$, equal in all channels.

The other system we consider are metallic carbon nanotubes (NT).
Here, spin-orbit coupling \cite{martino} arise predominantly
from the curvature of the tube surface, cf.\ Ref.\
\onlinecite{magarill}; in flat graphite layers it vanishes by
mirror symmetry. The kinetic energy dispersion may be taken as
$\epsilon_n^{\mbox{\tiny
NT}}(k)=\pm\{\epsilon_n+\vf[|\kf-k|\Theta(k)+|\kf+k|\Theta(-k)]\}$,
disregarding, for simplicity, parabolic parts of the dispersion
close to the subband bottoms at $\epsilon_n$. Here, $\kf$
denotes the Fermi momentum at zero doping and $\vf$ the Fermi
velocity. For every $n$ there exist two different flavor branches
$b=\pm 1$, depending on the magnitude $|k|{>\atop<}|\kf|$, of
opposite velocities at given sign of $k$. Adding $H^{\rm so}$
from Eq.~(\ref{hso}) yields differences
$k_{n{\tiny\uparrow}}-k_{n{\tiny\downarrow}}=
2\alpha(\ef-\epsilon_n+b\vf\kf)/(\vf^2-\alpha^2)$ at the Fermi
energy, and therefore Rashba precession lengths, depending now
on $n$ and, additionally, on $b$.

So far we considered independent electrons for QW and NT. It
has been shown recently that electron-electron interactions
influence and actually enhance Rashba precession in 2D
structures \cite{chenraikh} and in single transport channels
\cite{rashbawh}. To calculate Rashba lengths on the basis of
the multi channel TL model (\ref{mbose}) requires expressing
Eq.~(\ref{hso}) in Bose variables. As in single channels
\cite{rashbawh} $H_{\rm so}$ is found to be proportional to the
spin currents $J_{n,\sigma}$. The expressions
\begin{equation}\label{hsoqw}
H_{\rm so}^{\mbox{\tiny QW}}=-\alpha m_{\rm e}\sum_nv_nJ_{n,\sigma}
\end{equation}
and
\begin{equation}\label{hsont}
H_{\rm so}^{\mbox{\tiny NT}}=-\alpha\vf\sum_{n\atop\epsilon_n<|\ef|}
\frac{\ef-\epsilon_n+b\vf\kf}{\vf^2-\alpha^2}J_{n,\sigma}\;,
\end{equation}
however, differ in the two cases, QW and NT, as can be checked
in the limit of vanishing interaction, when
$(\bm{v}_{\mbox{\tiny N}\sigma})_{nn'}=(\bm{v}_{\mbox{\tiny
J}\sigma})_{nn'}=v_n\delta_{nn'}$ in Eq.~(\ref{mbose}) reproduces
the Rashba precession lengths
$2\pi/|k_{n{\tiny\uparrow}}-k_{n{\tiny\downarrow}}|$ in
individual QW or NT channels, respectively ($H_{\rm so}$ is a
single particle operator and does not depend on the Coulomb
interaction).

With Eqs.~(\ref{hsoqw}) or (\ref{hsont}) we are now in the position
to calculate
\begin{equation}\label{rp}
f(L)=\frac{1}{M}\sum_{nn'}\langle\bigl(\psi_{n\uparrow}^{}(L)+
\psi_{n\downarrow}^{}(L)\bigr)\bigl(\psi_{n'\uparrow}^{\dagger}(0)+
\psi_{n'\downarrow}^{\dagger}(0)\bigr)\rangle\;,
\end{equation}
employing the well known Boson representation for Fermi
operators $\psi_{ns}^{}(x)$ \cite{haldane}. In the spirit of
the Datta-Das setup \cite{dattadas} the quantity $f(L)$ can be
interpreted as the probability amplitude to measure an electron
at $x=L$ with its initial spin polarization parallel to the
$x$-axes, provided it was injected at $x=0$ with equal
probability amplitudes into all occupied channels. The result
can be represented as
\begin{equation}\label{fl}
f(L)=\sum_{nn'}g_{nn'}(L)\cos(\pi L/\lambda_n)\;.
\end{equation}
As indicated in the previous section the asymptotic power law
decay of the two point function
$g_{nn'}(x)\sim|x|^{-\beta_{nn'}}$ can be expressed through
the solutions of Eq.~(\ref{bogo2}) for {\bf R} and $\vec\vartheta$:
\begin{equation}\label{betann}
\beta_{nn'}=\delta_{nn'}+2\sum_j{\rm R}_{nj}{\rm
R}_{n'j}\sinh^2\vartheta_j\;.
\end{equation}
Equation~(\ref{betann}) generalizes the known result for the electron
Green function of single channels \cite{larkin,haldane}. By
virtue of conformal invariance the time dependence can be
deduced, yielding the open boundary exponent
\begin{equation}\label{betan}
\beta_n=-1+\sum_{n'}{\rm R}_{nn'}^2{\rm e}^{-2\vartheta_{n'}}
\end{equation}
for the density of states [Eq.~(\ref{exponent})].

Equation~(\ref{fl}) allows to read off the inverse precession lengths
\begin{equation}\label{lambdaqw}
\frac{1}{\lambda_n^{\mbox{\tiny QW}}}=\frac{\alpha m_{\rm e}}{\pi}\sum_{n'}
(\bm{v}_{\mbox{\tiny J}\sigma}^{-1})_{nn'}v_{n'}
\end{equation}
and
\begin{eqnarray}\label{lambdant}
\d\frac{1}{\lambda_n^{\mbox{\tiny NT}}}&=&\d\frac{\alpha\vf}{\pi}
\sum_{n'\atop\epsilon_{n'}<|\ef|}
(\bm{v}_{\mbox{\tiny J}\sigma}^{-1})_{nn'}\\
&&\hspace*{15mm}\d{}\times
\frac{\ef-\epsilon_{n'}+b\vf\kf}{\vf^2-\alpha^2}\;,\nonumber
\end{eqnarray}
after which spins reverse their polarization in channel $n$. We
see that the $\lambda_n$ are governed by the matrices
$\bm{v}_{\mbox{\tiny J}\sigma}$ of spin conductivities. Without
electron-electron interaction $(\bm{v}_{\mbox{\tiny
J}\sigma})_{nn'}=v_n\delta_{nn'}$ so that
$\lambda_n^{\mbox{\tiny QW}}=\lambda$ stay equal in all channels
of QW [Eq.~(\ref{lambdaqw})]. This would yield optimum transistor
operation as discussed before. Interactions, however, alter the
diagonal entries and, additionally, generate off-diagonal
entries in $\bm{v}_{\mbox{\tiny J}\sigma}$, describing the
coupling between channels. In general $\lambda_n^{\mbox{\tiny
QW}}$ become $n$-dependent so that different channels dephase.
Remarkably, as demonstrated now, interactions {\em reduce\/} spin
dephasing in the two lowest flavor modes of single walled NT
[Eq.~(\ref{lambdant})], compared to its magnitude in the absence
of interactions.

\section{Perturbative estimate}
Sufficiently weak interactions can justify the perturbative
estimate to $\bm{v}_{\mbox{\tiny J}\sigma}$. Crucial is the
magnitude of nonzero Fourier components, which even in metallic
NT are small compared to the Fermi velocity, since the (large)
$q=0$ component of the interaction does not affect the spin
sector and is properly accounted for on the RPA level. Imposing
SU(2) symmetry [naive low order perturbation theory violates
SU(2) invariance \cite{whlkahm}] results in
\begin{eqnarray}\label{perturbative}
&&\d (\bm{v}_{\mbox{\tiny
N}\sigma})_{nn'}=(\bm{v}_{\mbox{\tiny
J}\sigma})_{nn'}=[v_n-\{V_{nnnn}(2k_n)\nonumber\\
&&\d\mbox{}+\sum_{j\ne n}V_{njnj}(k_n+k_j)-V_{njnj}(k_n-k_j)\}/2\pi]
\delta_{nn'}\nonumber\\
&&\d\mbox{}-[V_{nn'nn'}(k_n-k_{n'})/2\pi](1-\delta_{nn'})\;.
\end{eqnarray}
Here, $k_n$ denotes the Fermi momentum of mode $n$ \cite{redistribute}.

\begin{figure}
\centerline{\raisebox{5mm}{\epsfxsize=0.5\columnwidth\epsffile{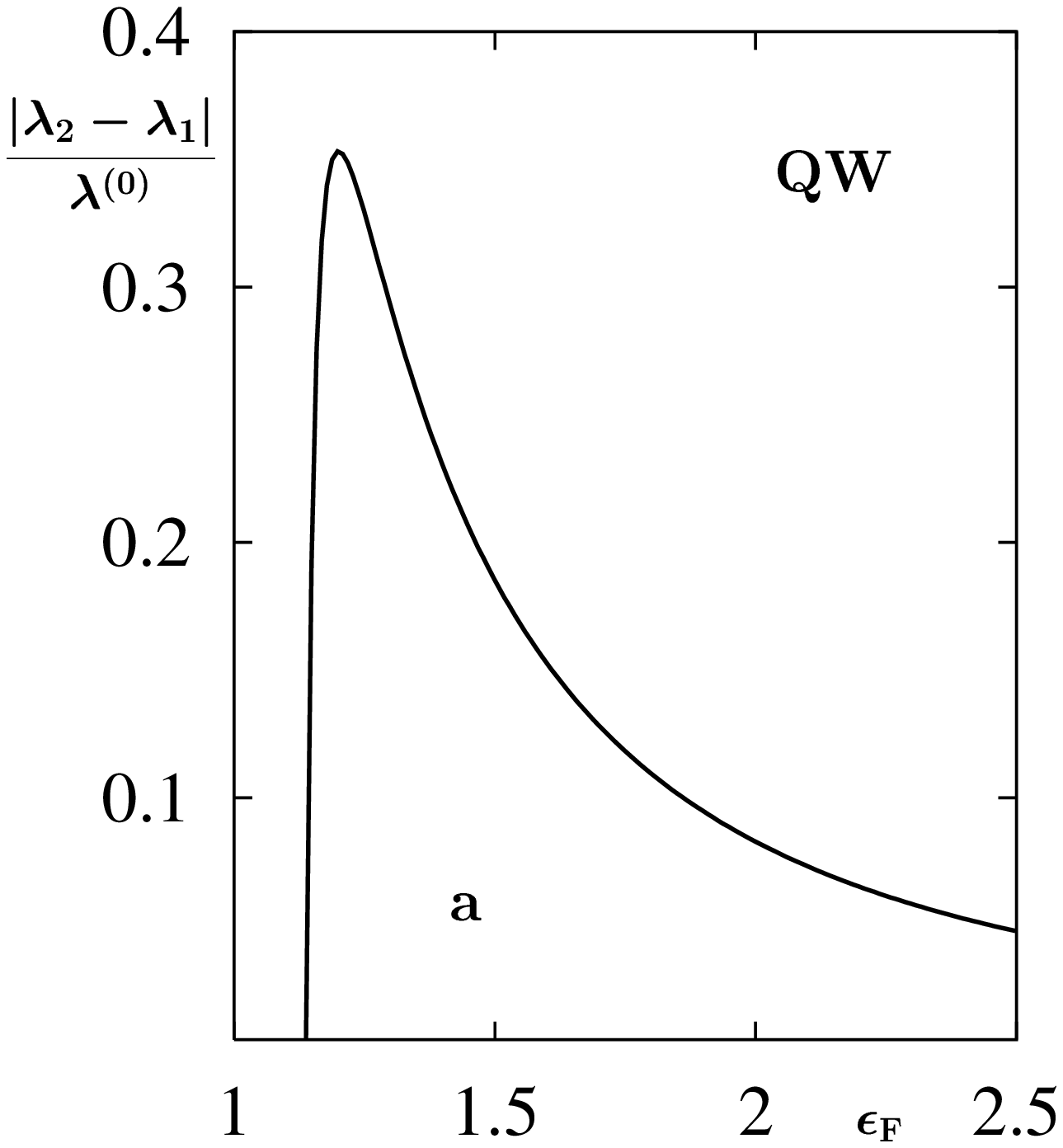}}
\epsfxsize=0.5\columnwidth\epsffile{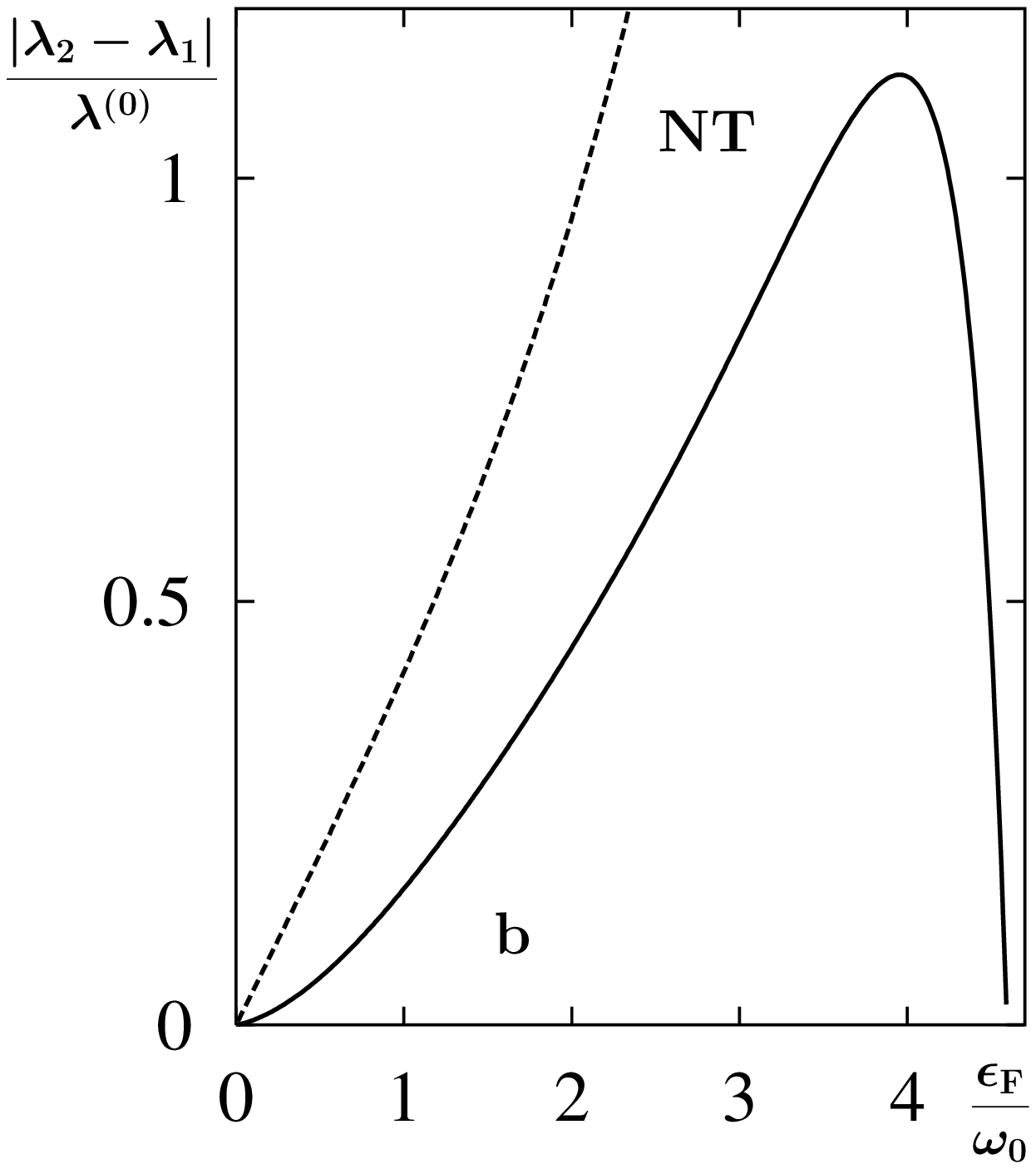}}
\caption[]{
Difference $|\lambda_2-\lambda_1|/\lambda^{(0)}$ versus Fermi
energy $\ef/\omega_0$ in units of the subband energy $\omega_0$
for two channels, cf.\ main text. In QW (a) interactions always
cause dephasing between both modes while in single walled NT (b)
dephasing between the two flavor modes can be significantly
suppressed at small doping compared to the noninteracting case
(dashed).}
\label{lamdr}
\end{figure}

Figure~\ref{lamdr}(a) shows the resulting difference
$|\lambda_2-\lambda_1|/\lambda^{(0)}$ of Rashba lengths relative
to this length $\lambda^{(0)}$ in the absence of interactions in
a two-channel QW of width $d=a_{\rm B}/\sqrt{2}$ ($a_{\rm B}$ is
the Bohr radius). This phase difference can exceed 30\% and
decreases only when $\ef$ and the carrier density increase so
that the interaction strength diminishes. We see that dephasing
never vanishes within the regime of validity of the
perturbational approach at not too small carrier densities in
the second subband, $\ef\gtrsim 1.15\omega_0$ [note that for
harmonic confining potential the third subband (not included
here) becomes occupied above $\ef/\omega_0>2$]. Experimentally,
it might be possible to disentangle two Rashba periods and thus,
by monitoring their dependence on the interaction strength
(carrier density), verify the predicted dephasing mechanism
when two channels are occupied. For the Datta Das transistor
based on semiconductor QWs this result clearly suggests to use
only the lowest subband.

In Figure~\ref{lamdr}(b) the difference of Rashba lengths of the
two lowest (degenerate) flavor modes of metallic single wall arm
chair ($m,m$) NT is seen for $m=5$ [subband energies
$\epsilon_1=\epsilon_2=0$, the Fermi momenta $k_{\{{1\atop
2}\}}=m/\sqrt{3}r\pm\ef/\vf$, $r$ is the tube radius, and
$e^2/\ve\vf\approx 2.7$ \cite{eg}], relative to the splitting
$\lambda^{(0)}$ at zero doping. Dephasing can shown to be
suppressed logarithmically
$\sim\ef/[c_1-c_2\ln(\ef/\omega_0)]+{\cal O}(\ef^3)$ at small
doping ($c_1$ and $c_2$ depend on the tube radius and $c_2$
vanishes with the interaction), compared to the interaction free
case (dashed) where $c_1=m/2$. This suggests the use of single
walled NT close to the neutrality point \cite{massgap} for
coherent Rashba precession along both flavor channels
\cite{multiwall}.

\section{Summary}
We have calculated Rashba spin precession lengths $\lambda_n$ in
multimode quantum wires, accounting for electron-electron
interactions, using the Bosonization method. To this end we
have developed an in principle exact description for the
TL phase of coupled quantum channels that allows the inclusion
spin. Generalized charge and spin compressibilities and
conductivities are identified as the key parameters to determine
the power law exponents and to establish quantitative contact
with the underlying interacting electron model. In
semiconducting wires, characterized by a parabolic kinetic
energy dispersion, we find that the $\lambda_n$ become
$n$-dependent, giving rise to doubts whether multichannel
systems can be used as active part of the Datta-Das transistor.
This result could explain the up to date lack of successful
transistor operation and clearly suggests to use single channel
quantum wires. In metallic single walled carbon nano\-tubes, on
the other hand, we find that dephasing between the two flavor
modes, arising due to the linear kinetic energy dispersion, is
suppressed by the electron-electron interaction, particularly at
small doping, which could make these systems interesting for
coherent spin transport.
\vspace{5mm}

I thank Hermann Grabert for many useful conversations. This
work has been partly supported by the DFG under SFB~276.

\end{multicols}%

\begin{thebibliography}{10}
\bibitem{conductqu} This property is related with conductance
 quantization of ballistic wires.
\bibitem{redistribute} Interactions may redistribute
 carriers between the channels for the ground state; here,
 $2k_n/\pi$ is the carrier density as measured in channel $n$
 {\em after\/} this relaxation has taken place. Additional
 redistribution due to $H_{\rm so}$ is negligible.
\bibitem{multiwall} When doping occupies excited angular
 momentum states, as often in multi-walled NT, we expect similar
 dephasing as in quantum wires.
\end{thebibliography}
\end{document}